\documentclass[aip,jap,reprint,numerical]{revtex4-1}
\usepackage[colorlinks=true,linkcolor=blue,citecolor=blue,urlcolor=blue]{hyperref}
\usepackage{amsmath}
\usepackage{amssymb}
\usepackage[pdftex]{graphicx}
\usepackage{epstopdf} 
\usepackage{graphicx}
\usepackage{grffile}
\usepackage{sidecap}
\sidecaptionvpos{figure}{t}
\usepackage{dcolumn}
\usepackage{bm}
\begin{document}
\title{Large anisotropic spin relaxation time of exciton bound to donor states in triple quantum wells}
\author{S. Ullah}
\author{G. M. Gusev}
\affiliation {Instituto de F\'{i}sica, Universidade de S\~{a}o Paulo, Caixa Postal 66318 - CEP 05315-970, S\~{a}o Paulo, SP, Brazil}
\author{A. K. Bakarov}
\affiliation{Institute of Semiconductor Physics and Novosibirsk State University, Novosibirsk 630090, Russia}
\author{F. G. G. Hernandez}
 \email[Corresponding author.\\ Electronic address: ]{felixggh@if.usp.br}
\affiliation {Instituto de F\'{i}sica, Universidade de S\~{a}o Paulo, Caixa Postal 66318 - CEP 05315-970, S\~{a}o Paulo, SP, Brazil}
\date{\today}
\begin{abstract}
We have studied the spin dynamics of a dense two-dimensional electron gas confined in a GaAs/AlGaAs triple quantum well by using time-resolved Kerr rotation and resonant spin amplification. Strong anisotropy of the spin relaxation time up to a factor of 10 was found between the electron spins oriented in-plane and out-of-plane when the excitation energy is tuned to an exciton bound to neutral donor transition. We model this anisotropy using an internal magnetic field and the inhomogeneity of the electron $g$-factor. The data analysis allows us to determine the direction and magnitude of this internal field in the range of a few mT for our studied structure, which decreases with the sample temperature and optical power. The dependence of the anisotropic spin relaxation was directly measured as a function of several experimental parameters: excitation wavelength, sample temperature, pump-probe time delay, and pump power.
\end{abstract}
\maketitle 
\section{Introduction}
The study of spin dephasing and spin relaxation processes for carriers in two-dimensional electron gases (2DEGs) confined in semiconductor quantum wells (QWs) is a necessary step for building practical spintronics devices.\cite{Datta1990,Zhukov2006,Zhukov2007} The spin-orbit interaction provides a way to electrically control the carrier spins due to the energy splitting induced by structure inversion asymmetry (Rashba)\cite{Rashba1984} or bulk inversion asymmetry (Dresselhaus)\cite{Dresselhaus1955}. On the other hand, the carrier spins will randomly precess about a momentum-dependent effective magnetic field opening a channel for spin relaxation via so-called Dyakonov-Perel (DP) mechanism.\cite{Dyakonov1972}\par
The anisotropy of spin relaxation in semiconductor nanostructures has been widely studied in bulk\cite{Lee2015} as well as in QWs grown along different orientations.\cite{Averkiev1999,Averkiev2006,Dohrmann2004,Morita2005,Liu2007,Stich2007,Korn2008,Zhao2014} In zinc-blende heterostructures with equal strength of the Rashba and Dresselhaus spin-orbit fields, an in-plane spin relaxation anisotropy was theoretically predicted \cite{Averkiev1999} and later experimentally confirmed by using Hanle effect.\cite{Averkiev2006} In that study, different spin relaxation times for the spins oriented along [110],[1$ \overline{1} $0] and [001] directions were evaluated. A number of experimental groups have extensively investigated the in-plane spin relaxation anisotropy. However, only a few have focused on the anisotropy between the relaxation times for the electron spins oriented parallel ($\tau_{z}$) and perpendicular ($\tau_{y}$) to the quantum well growth direction. For QWs grown along the z$\Vert$[001] axis, it was theoretically reported that the spin components of the resident carriers relax at different rates leading to strong anisotropy.\cite{Glazov2008} Later, a similar behavior was experimentally shown for [110] oriented GaAs QWs using the resonant spin amplification technique.\cite{Griesbeck2012}\par
It has been observed that the presence of anisotropic spin relaxation affects the relative amplitudes of the resonant spin amplification (RSA) peaks.\cite{Glazov2008,Griesbeck2012,Yugova2012} In the case of isotropic spin relaxation, all peaks at different fields have the same amplitude and the spin components of carriers oriented along $z$ and $y$ directions relax with the same rate. For anisotropic spin relaxation, the amplitude of the peak corresponding to zero external magnetic field differs from the peaks at finite field $B_{ext}\parallel x$.\cite{Glazov2008,Yugova2012}\par
In the present work, we are interested in the study of anisotropic relaxation for multiple quantum well systems. Similar samples have allowed the discovery of interesting phenomena, such as intrinsic spin Hall effect\cite{Felix2013}, collapse of the integer quantum Hall effect\cite{Boebinger1990,MacDonald1990}, long-lived spin coherence\cite{Saeed2016}, spontaneous interlayer phase coherence\cite{Eisenstein1992,Spielman2000,Eisenstein2003}, control and drift current induced spin polarization\cite{Felix2014,Felix2016}, and excitonic Bose condensation.\cite{Kellogg2004,Tutuc2004,Eisenstein2004}. Here, we report on the experimental observation of large spin relaxation anisotropy in triple quantum wells. We obtained long relaxation times for the spins oriented along the growth direction and fast relaxation times for the spins oriented in-plane when the excitation is tuned to an exciton bound to a neutral donor transition. We model the anisotropy in terms of an effective in-plane magnetic field and analyzed how the experimental parameters influence the spin relaxation anisotropy in our structure.\par
\begin{figure*}
\includegraphics[width=\textwidth]{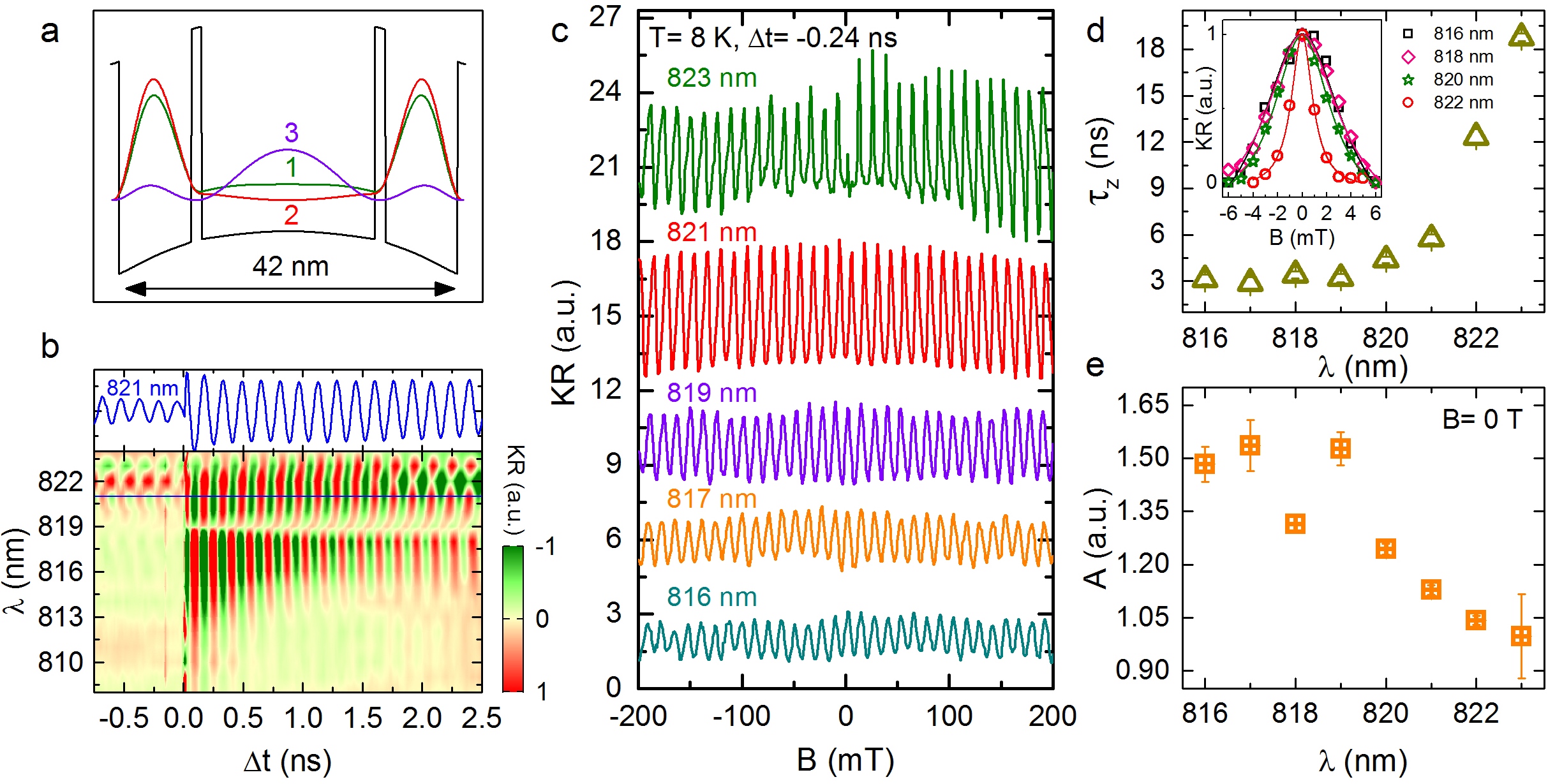}
\caption{(a) Band diagram and charge density for the studied TQW sample. (b) Pump-probe delay scans of the KR signal measured for different wavelengths at $ B $ = 1 T and $ T $ = 8 K, where the solid line highlights the time evolution of spin dynamics at $\lambda$ = 821 nm as shown on the top of contour plot. (c) $ B_{ext} $ scan of KR for different pump-probe wavelengths at fixed pump-probe delay. (d) Spin relaxation time along $z$ and (e) amplitude of zero-field peak as a function of pump-probe wavelengths extracted from (c). The inset shows a Lorentzian fit to the zero-field resonance peaks.}
\label{fig:wide}
\end{figure*}
\section{Materials and Experiment}
We studied a dense two-dimensional electron gas confined in GaAs/AlGaAs triple quantum well (TQW) grown in the [001] direction by molecular beam epitaxy.\cite{Wiedmann2009} The central well width is about 22 nm, and both side wells have equal widths of 10 nm each separated by 2-nm-thick $Al_{0.3}Ga_{0.7}As$ barriers. The central well width has been kept wider than the lateral wells in order to be populated because the electron density tends to concentrate mostly in the side wells as result of electron repulsion and confinement. The sample is symmetrically $\delta$-doped with total electron sheet density of $n_{s} = 7 \times 10^{11}$ cm$^{-2}$ and low-temperature mobility $\mu = 4 \times 10^{5}$ cm$^2$/Vs. The interlayer coupling results in the formation of three subbands with subband separations of $\Delta_{12}= 1.0$ meV, $\Delta_{23}= 2.4$ meV and $\Delta_{13}= 3.4$ meV. The calculated TQW band structure and the subband charge density are depicted in Fig. \ref{fig:wide}(a), where the black lines correspond to the potential profile while the colored lines show the first (green), second (red) and third (purple) occupied subbands.\par
We used pump-probe Kerr rotation (KR) and RSA to study the coherent spin dynamics of the QW 2DEG. The light source was a mode-locked Ti-sapphire laser, with repetition rate of $f_{rep}$ = 76 MHz ($t_{rep}$ = 13.2 ns), delivering 100 fs pulses. Spin polarization along the QW growth direction was optically created by exciting the sample with a circularly polarized pump and detected with time-delayed probe pulses. The time delay $\Delta t$ between pump and probe pulses was varied by a mechanical delay line. For lock-in detection, the circular polarization of the pump beam was modulated at 50 kHz by a photo-elastic modulator. The probe beam was linearly polarized, and the polarization rotation of the reflected beam was detected with a balanced bridge using coupled photodiodes. The laser energy was tuned to obtain the largest time-resolved Kerr rotation (TRKR) signal in the sample. The sample was mounted in a cryostat for measurements under a transverse magnetic field along x (Voigt geometry), which forced the spins to precess around the field and allows us to monitor its relaxation time in the zy plane.
\section{Results and discussion}
\subsubsection{\label{sec:Wave}Wavelength dependence of spin dynamics}
TRKR traces measured for different pump-probe wavelengths at a pump power of 1 mW, $ B $ = 1 T and $ T $ = 8 K are shown in Fig. \ref{fig:wide}(b). For this pump density of 50 W/cm$^2$, the photo-excited carrier density is comparable to the 2DEG density. The striped pattern corresponds to the precession of coherently excited electron spins about the in-plane magnetic field. As reported,\cite{Zhukov2007,Saeed2016,Zhukov2012,Zhukov2014} long-lived spin coherence can be identified by the carrier precession at negative $ \Delta t $ just before the pump pulse arrival. One can clearly see those negative delay oscillations in the wavelength range from 817 nm to 823 nm, which are more pronounced at higher wavelengths as shown on the top of contour plot for clarity. After the pump pulse arrival at $ \Delta t $ = 0, the generated spin polarization shows a very weak decay over the time window of 2.5 ns. Increasing the wavelength from 814 nm up to 818 nm also leads to an increase in the precession frequency due to a variation of the g-factor with a sudden change visible at 819 nm.\par
TRKR profile in Fig. \ref{fig:wide}(b) shows two distinct regions with pronounced oscillations corresponding to $ \lambda $ = 817 nm and 820 nm with the $ g $-factors of 0.461 and 0.452 (absolute values), respectively. Those regions are in agreement with the two distinct lines observed in the magneto-photoluminescence study performed on the same sample.\cite{Pusep2014} The line corresponding to high wavelengths was attributed to the exciton bound to a neutral donor (DX center), while the one at low wavelength was related to the direct recombination between the states confined in the conduction and valence band. Also, it was argued that the interband recombination of the electrons located on the Landau levels mediated by DX centers contribute to the line related to the emission caused by DX centers.\cite{Pusep2014} \par
Figure \ref{fig:wide}(c) displays the RSA signals measured in the wavelength range where strong negative delay oscillations were observed. The time delay between pump and probe was adjusted such that the probe pulse arrives 0.24 ns before the subsequent pump pulse, and the Kerr signal was recorded by sweeping the magnetic field over the range from -200 mT to 200 mT. The RSA signal consists of a series of sharp peaks with spacing of $\Delta B = (h/(g \mu_Bt_{rep}))$ where $ h $ is Planck's constant, $ \mu_{B} $ is Bohr magneton, $ g $ is the electron g-factor, and $ t_{rep} $ is the laser repetition period.\cite{Kikkawa1998} Those resonance peaks correspond to the carrier spin precession frequencies which are commensurable with the pump pulse repetition period. To get information on the spin relaxation times, the resonance peaks were fitted with a Lorentzian model:\cite{Kikkawa1998,Felix2014} 
\begin{equation}
\Theta_{K}(B) = A/[(\omega_{L}\tau_{s})^{2}+1]
\label{eq:Lorentz}
\end{equation}
with full width at half maximum $B_{1/2}=\hbar/(g\mu_{B}\tau_{s})$, where  $ A $ is KR amplitude, $ \tau_{s} = [2\tau_{y}\tau_{z} /(\tau_{y} + \tau_{z})] $\cite{Yugova2012} and $ \omega_L= g\mu_{B}B/\hbar$ is the Larmor precession frequency. $ \tau_{z} $ is the carrier spin relaxation in the absence of magnetic field while the side peaks at $ B\neq0 $ reveal the spin relaxation time $ \tau_{s} $. The inset in Fig. \ref{fig:wide}(d) shows the comparison of four (normalized) zero-field peaks with fitted Lorentzian curves for several wavelengths.  The extracted values of $\tau_{z}$ and amplitude are plotted in Fig. \ref{fig:wide}(d) and (e) as a function of wavelength.\par
Changing the pump-probe energy about 5.6 meV ($ \simeq 2\Delta_{23} $), by increasing the wavelength from 816 nm up to 819 nm, has almost no effect on the spin relaxation time that remains constant around 3 ns. Further increasing the wavelength causes an rapid variation in the dephasing time and yields 18.8 ns for $ \lambda $= 823 nm. In contrast, the amplitude of zero-field peak reduces sharply with higher wavelengths. We see that the RSA peaks centered around zero magnetic fields are suppressed in comparison with the neighbouring peaks for higher wavelengths. \par
It has been demonstrated, in a energy-resolved study of spin relaxation for n-type bulk GaAs doped beyond metal-insulator transition, that a strong variation of relaxation time occurs when occupying the conduction and donor band states.\citep{Schreiber2007} The longest lifetime, exceeding 100 ns, was found for the delocalized donor band states. Furthermore, in n-type bulk ZnO, it has been reported that for donor-bound electrons the spin relaxation is strongly anisotropic.\citep{Lee2015} It was claimed that the defects present in the sample due to impurities, induce bound exciton in the vicinity of donor states which further contribute to the spin signal. This contribution to the bound electron spins at donor sites is more predominant at low temperatures and leads to relaxation anisotropy. In our structure such effect is also stronger at the DX transition energy, therefore, we limited our study to $ \lambda $ = 823 nm.\par
\begin{figure}[t]
\includegraphics[width=0.75\columnwidth]{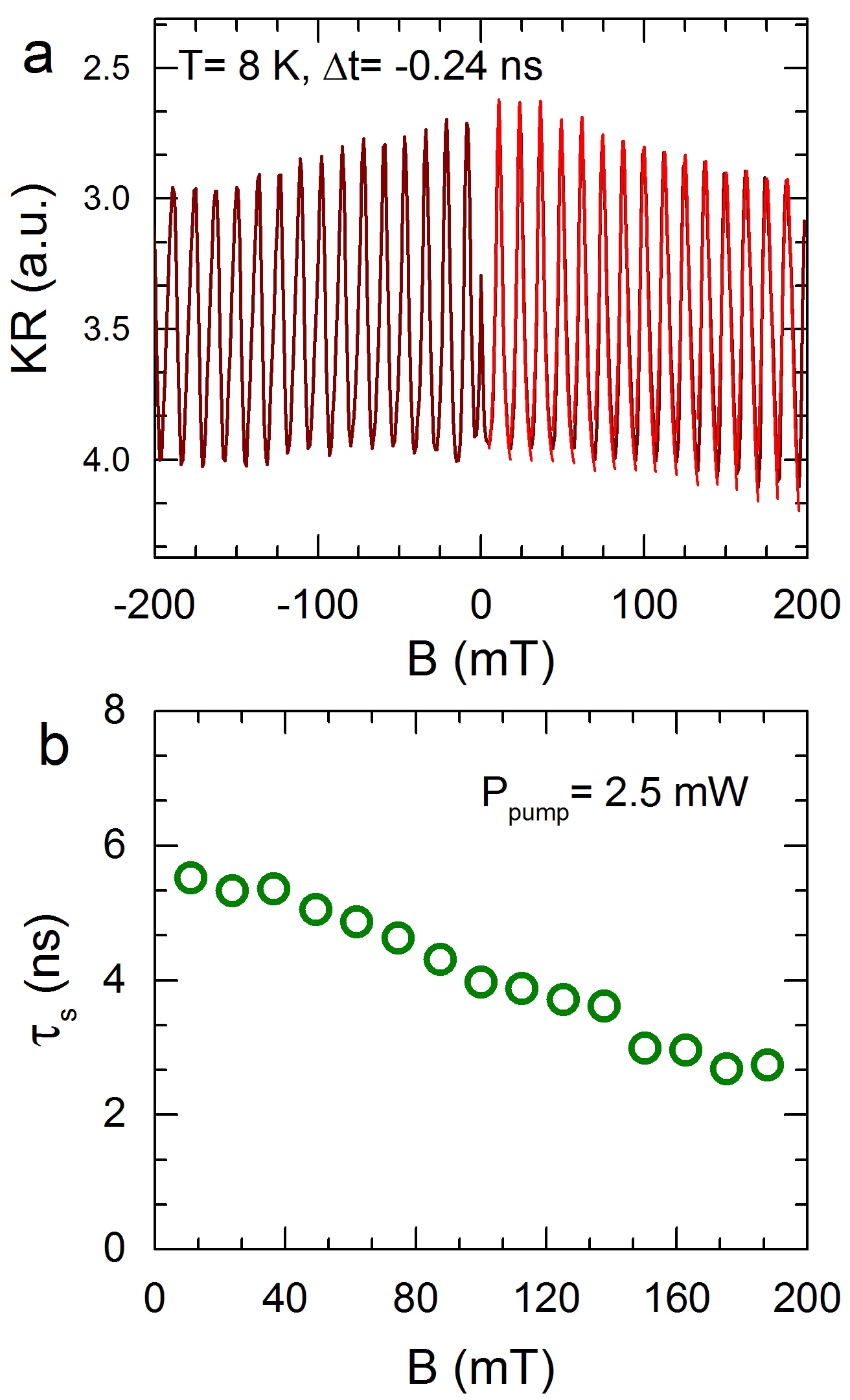}
\caption{(a) $ B_{ext} $ scan of KR signal at $ \Delta t $ = -0.24 ns and P$_{pump}$ = 2.5 mW. The red line is Lorentzian fit from where $ \tau_{s} $ is extracted and plotted in (b).}
\label{fig:RSAvsBext}
\end{figure}

\begin{SCfigure*}
\includegraphics[width=0.77\textwidth]{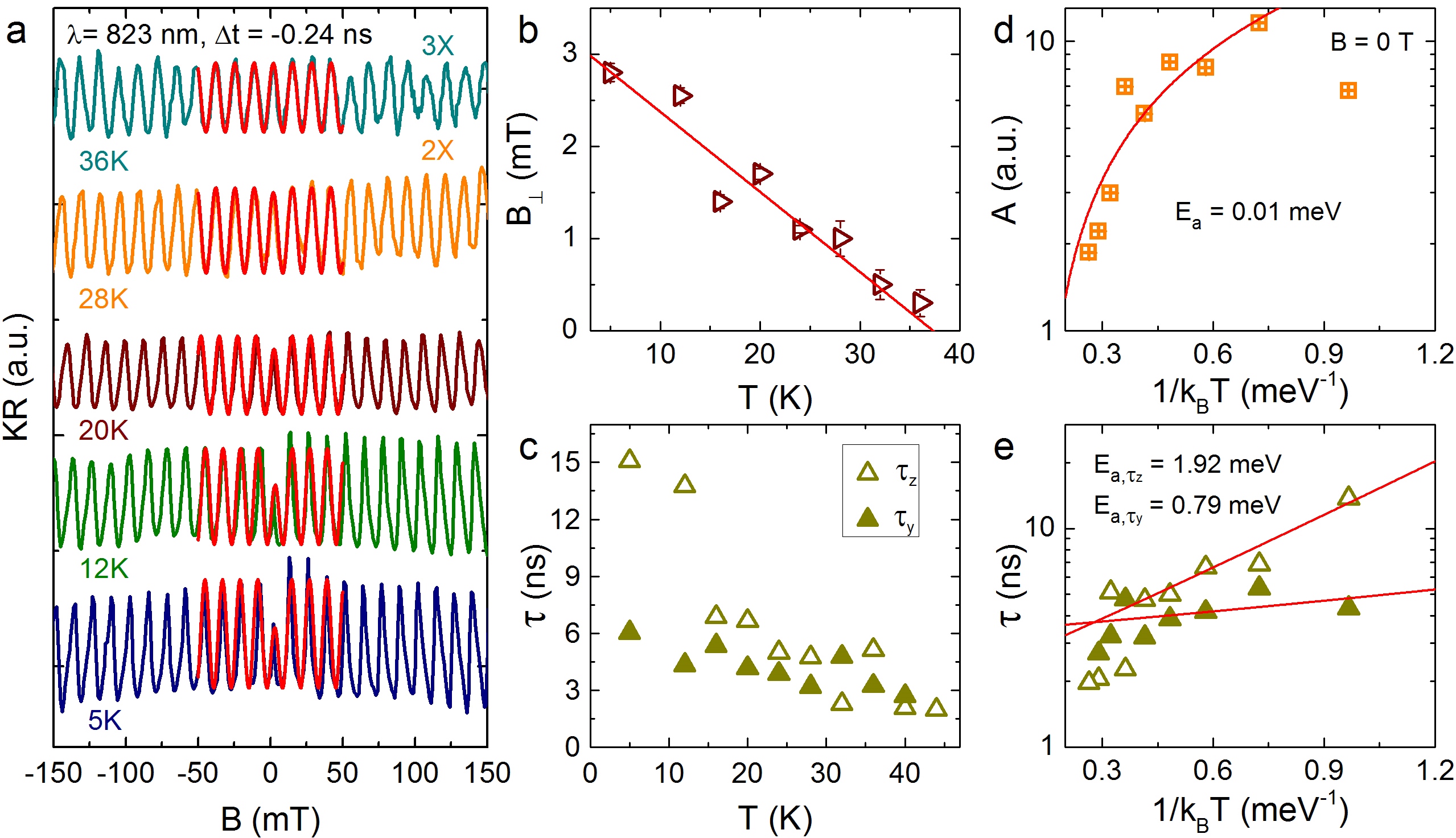}
\caption{ Temperature influence on resonant spin amplification (a) Temperature dependence of RSA signal. The red line is a fit to the data with Eq. \ref{eq:Bso}. (b) Linear dependence of $ B_{\perp} $ as a function of temperature with slope $-8\times$10$^{-2}$mT/K. (c) $\tau_{z}$ (open triangles), $\tau_{y}$ (solid triangles) and (d) amplitude (squares) extracted from Lorentzian fit. (e) Relaxation time for the y and z components of spin relaxation time as a function of reciprocal thermal energy. Fittings in d) and e) are Arrhenius-like functions giving the activation energies E$_a$.}
\label{fig:tempdependence}
\end{SCfigure*}

\subsubsection{Effect of inhomogeneous broadening}
Figure \ref{fig:RSAvsBext}(a) shows the magnetic field scan at pump power of 2.5 mW and $ \Delta t $ = -0.24 ns. For the analysis of the external magnetic field influence on the spin relaxation time, every single resonance peak for positive field was fitted to the Lorentzian model as shown by the solid red curves on the top of experimental data in panel (a). The extracted values of $ \tau_{s} $ are presented in Fig. \ref{fig:RSAvsBext}(b) showing a strong reduction with increasing field. The broadening of RSA peaks and the reduction of their amplitude with the higher magnetic field is caused by the inhomogeneity of the $ g $-factor.\cite{Glazov2008}
To separate the effect of the $ g $-factor inhomogeneity with growing magnetic field from other relaxation components, $ \tau_{s} $ was evaluated from the first peak next to the zero-field peak in the following discussion.

\subsubsection{Temperature influence on spin relaxation anisotropy\label{sec:Temp}}
We turn now to the analysis of the effect of the sample temperature on the evolution of the carriers spin dynamics. Figure \ref{fig:tempdependence}(a) shows the RSA traces recorded at different temperatures while keeping the time delay between pump and probe fixed. The magnetic field was scanned while slowly heating up the sample, starting from $ T $ = 5 K up to higher temperatures until the RSA signal vanished. The observed RSA pattern is well described by \cite{Norman2014}
\begin{equation}
\Theta_{K}(B) = A cos \left(\frac{g \mu_{B} \Delta t}{\hbar}\sqrt{(B_{ext} + B_{\parallel})^{2} + B_{\perp}^{2}}\right)
\label{eq:Bso}
\end{equation}
where $ \hbar $ is the reduced Planck's constant and $ B_{\perp} $ and $ B_{\parallel} $ are the components of spin-orbit field perpendicular and parallel to the external magnetic field. For clarity of presentation, the fitting using Eq. \ref{eq:Bso} is shown for reduced range of the RSA pattern from -50 mT to +50 mT. The extracted value of $ B_{\parallel} $ was negligible in the studied structure and is omitted here. The magnitude of $ B_{\perp} $ is plotted in Fig. \ref{fig:tempdependence}(b) showing a  linear decrease with increasing temperature in the 5-36 K range. \par
The fitted values of $ \tau_{z} $, $ \tau_{y} $ and amplitude using equation \ref{eq:Lorentz} are depicted in Fig. \ref{fig:tempdependence}(c) and (d). The relaxation times for both spin orientations decrease monotonically with the temperature and are qualitatively similar above 15 K. The relaxation time anisotropy is stronger at low temperatures and decreases with temperature reaching a constant value above 20 K. The relaxation times drop to $ \sim $ 3 ns at 40 K. The temperature dependence of the relaxation time may show the characteristic DP spin relaxation mechanism in QWs.\cite{Dyakonov1971,Dyakonov1972} Dyakonov and Kachorovskii \cite{Dyakonov1986,Munoz1995} found an inverse dependence for the spin relaxation time with temperature [see Fig. \ref{fig:tempdependence}(e)] due to the variation of the electron momentum relaxation time. With the occupation of higher momentum states, the DP relaxation mechanism becomes more efficient at elevated temperature. The amplitude of the zero-field peaks and the relaxation times (Fig. \ref{fig:tempdependence}(d) and (e)) were fitted using Arrhenius-like functions giving thermal activation energies E$_a$ below 2 meV. 

\subsubsection{Pump-probe time delay dependence of spin relaxation anisotropy}
\begin{figure}[t]
\includegraphics[width=0.95\columnwidth]{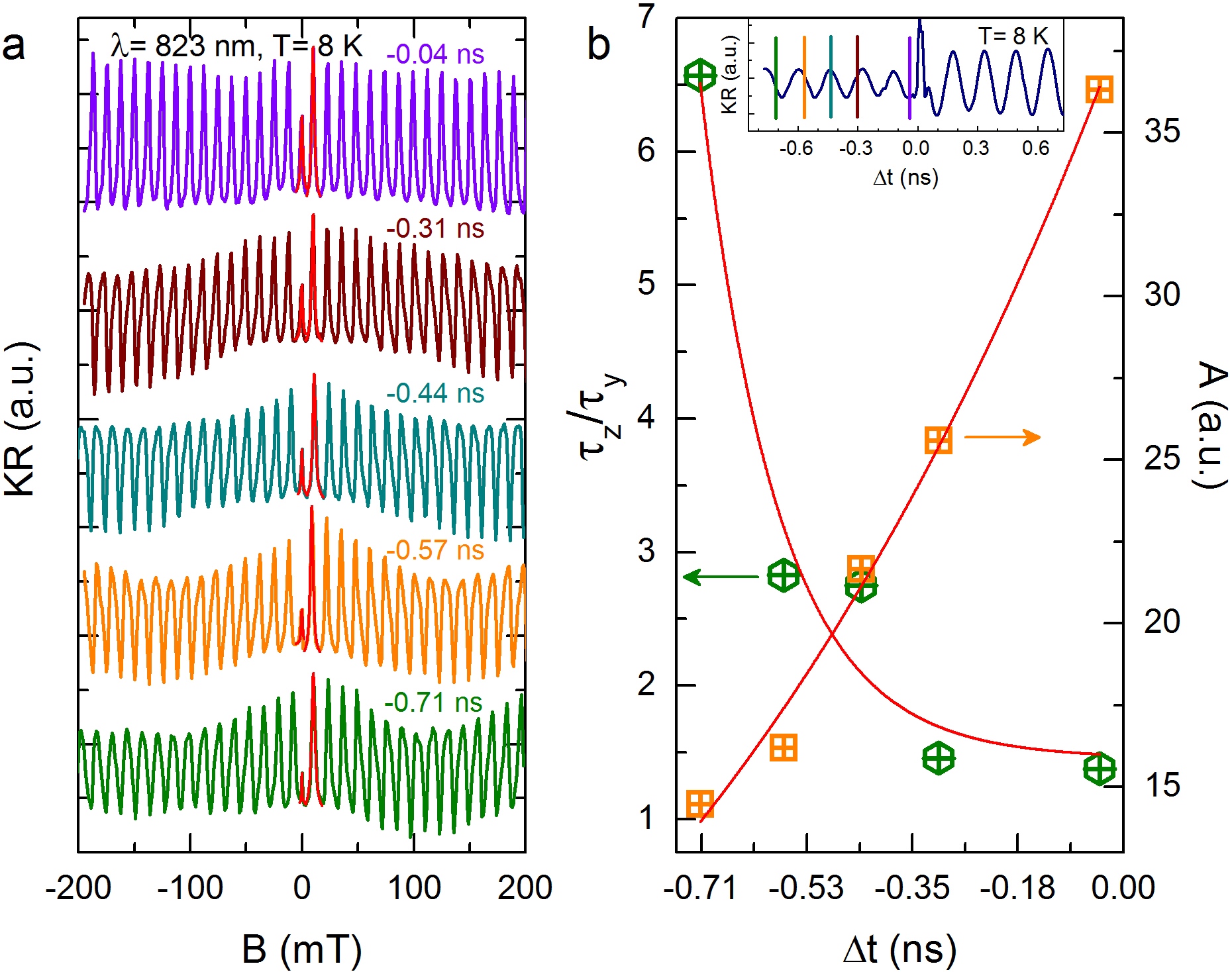}
\caption{Pump-probe delay dependence of resonant spin amplification (a) $ B_{ext} $ scan of KR signal at $ T $ = 8 K, for different time delays with pump/probe power of 1 mW/300 $ \mu $W. (b) Spin relaxation anisotropy and amplitude dependence on $\Delta t$. Inset Fig. (b) shows the time evolution of KR measured at 823 nm indicating the negative time delays for which RSA traces were recorded [lines of same color of the curves in (a)].}
\label{fig:delaydependence}
\end{figure}

In this section, we investigate the effect of KR signal at the different pump-probe delays. Figure \ref{fig:delaydependence}(a) displays the RSA signals measured with pump/probe power of 1mW/300$\mu$W. As function of the time delay, the shape of RSA signal differs from each other due to different phases in the spin precession.\cite{Yugova2012} Also, the resonances change the pointing direction because the spin resonance has precessed before the measurement during the delay time.\cite{Kikkawa1998} The spin relaxation anisotropy $ \tau_{z}/\tau_{y} $ and amplitude of the peaks corresponding to zero magnetic field obtained from the fit with eq. \ref{eq:Lorentz} are given in Fig. \ref{fig:delaydependence}(b). Increasing time delay between pump and probe causes the broadening of RSA peaks\cite{Saeed2016} and, as a result, the relaxation time and hence the anisotropy decreases. Inset in Fig. \ref{fig:delaydependence}(b) shows the time evolution of KR signal measured at $ B $ = 1 T. The lines at negative delay correspond to the delay points at which RSA traces were measured, where the line colors match the data presented in (a). $ \tau_{z}/\tau_{y} $ decreases from 6.6 at -0.71 ns down to 1.38 at -0.04 ns following an exponential decay (solid line). However, in contrast, the amplitude of the zero-field resonance peak increases exponentially for larger delay.
\subsubsection{Optical power dependence of spin relaxation anisotropy}
\begin{figure}[t]
\includegraphics[width=0.95\columnwidth]{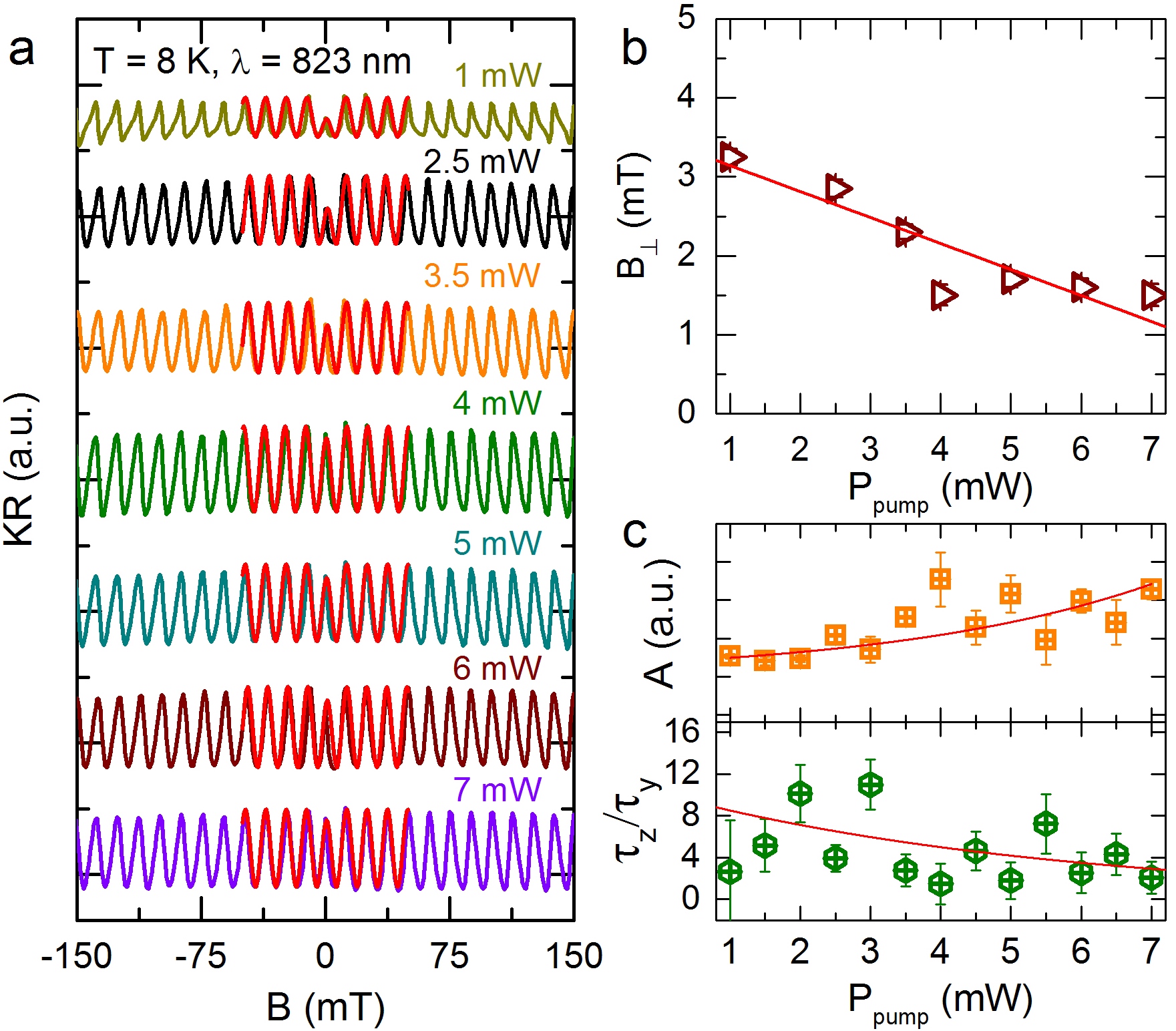}
\caption{Optical power dependence of resonant spin amplification (a) RSA signal for different optical power. The data have been shifted for clarity of presentation. The red curves on the top of experimental data are fit to Eq. \ref{eq:Bso} from where the in-plane internal magnetic field was extracted and plotted in (b). The linear fitting has a slope -0.3 mT/mW. (c) $ \tau_{z}/\tau_{y} $ and KR amplitude dependence on the optical power, where the solid lines are exponential fittings. $ T $ = 8 K, $ \Delta t $ = -0.24 ns. }
\label{fig:opticalpower}
\end{figure}

Finally, we report on the optical power influence on the spin relaxation anisotropy. Fig. \ref{fig:opticalpower} (a) presents the RSA signals measured in a range of pump power from 1 to 7 mW (corresponding to 50-350 W/cm$^2$) fitted to Eq. \ref{eq:Bso} in the same range of magnetic field of section \ref{sec:Temp}. The fitting result is displayed in Fig. \ref{fig:opticalpower} (b). As a function of pump power the $ B_{\perp} $ follows a linear decrease. From the Lorentzian fit to the data, $ \tau_{z}/\tau_{y} $ and signal amplitude were obtained and plotted as a function of pump power in Fig. \ref{fig:opticalpower} (c). Increasing the optical power causes the reduction of spin relaxation anisotropy. This reduction of $ \tau_{z}/\tau_{y} $ with high pump power can be correlated with the effect of heating of 2DEG and QW electron delocalization due to the interaction with photogenerated carriers \citep{Astakhov2008}. For 2DEGs confined in (001) GaAs/AlGaAs heterostructure, a similar decrease of an in-plane spin relaxation anisotropy with high pump power was attributed to the competition of spin-orbit interactions.\cite{Liu2007} The decrease of the spin relaxation time with high pump power is consistent with the previously reported data on the (001) oriented GaAs/AlGaAs double quantum wells.\cite{Saeed2016} The amplitude of the peaks at zero-field increase exponentially with high pump power.
\section{Conclusions}
In conclusion, we performed a detailed experimental study of the spin relaxation in a two-dimensional electron gas by TRKR and RSA. A strong anisotropy of the spin relaxation for the in- and out-of-plane spin orientations was observed when the excitation energy is tuned to an exciton bounded to a donor transition. The relaxation time for the spin oriented along the growth direction [001] is larger than the in-plane relaxation time. We model this anisotropy to the presence of in-plane internal field and to the $ g $-factor spread within the measured ensemble. The degree of anisotropy shows a strong dependence on the temperature and pump power with complete quenching of the anisotropy.

\section*{Acknowledgments}
The support of this work by Grants No. 2009/15007-5, No. 2013/03450-7, No. 2014/25981-7 and No. 2015/16191-5 of the S\~{a}o Paulo Research Foundation (FAPESP) is acknowledged. S.U. gratefully acknowledge TWAS/CNPq for the financial support. All measurements were done in the LNMS at DFMT-IFUSP.
\bibliographystyle{apsrev}
\bibliography{Ref}
\end{document}